\documentclass[secnumarabic,amssymb, nobibnotes, aps, prl]{revtex4-1}

\usepackage{graphicx}
\usepackage{gensymb}
\usepackage{color}

\begin{document}

\title{Enhanced thermal conductivity in percolating nanocomposites: a molecular dynamics investigation}%

\author{Konstantinos Termentzidis$^{1}$}
\author{Valentina M. Giordano$^{2}$}
\author{Maria Katsikini, Eleni C. Paloura$^{3}$}
\author{Gilles Pernot, David Lacroix,$^{4}$}
\author{Ioannis Karakostas$^{3,4}$}
\author{Joseph Kioseoglou$^{4}$}
\email{Konstantinos.Termentzidis@insa-lyon.fr}

\affiliation{
$^{1}$ Univ Lyon, CNRS, INSA-Lyon, Universit$\acute{e}$ Claude Bernard Lyon 1, CETHIL UMR5008, F-69621, Villeurbanne, France\\
$^{2}$ Institute of Light and Matter, UMR5306 Universit\'{e} Lyon 1-CNRS, Universit\'{e} de Lyon, 69622 Villeurbanne Cedex, France\\
$^{3}$ Universit$\acute{e}$ de Lorraine, LEMTA UMR 7563, 54504 Vandoeuvre les Nancy, France\\
$^{4}$ Universit\'{e} de Lorraine, LEMTA, CNRS-UMR7563, BP 70239, 54506 Vand\oe uvre Cedex, France.}

\date{10 October, 2018}%

\begin{abstract}

In this work we present a molecular dynamics investigation of thermal transport in a silica-gallium nitride nanocomposite. A surprising enhancement of the thermal conductivity for crystalline volume fractions larger than 5\% is found, which cannot be predicted by an effective medium approach, not even including percolation effects, the model systematically leading to an underestimation of the effective thermal conductivity. The behavior can instead be reproduced if an effective volume fraction twice larger than the real one is assumed, which translates in a percolation effect surprisingly stronger than the usual one. Such scenario can be understood in terms of a phonon tunneling between inclusions, enhanced by the iso-orientation of all particles. Indeed, if a misorientation is introduced, the thermal conductivity strongly decreases. We also show that a percolating nanocomposite clearly stand in a different position than other nanocomopsites, where thermal transport is domimnated by the interface scattering, and where parameters such as the interface density play a major role, differently from our case.
\end{abstract}

\maketitle

\section{Introduction}

Heterogeneous materials, made of the non alloying mixing of two different materials or phases, have recently come at the forefront of materials research thanks to the resulting novel physical properties, due partly to the combination of different properties from the constituent materials, and partly to the formation of interfaces between them. These latter become dominant in determining the functional properties of the novel material, when the mixing takes place at the nanoscale. Such materials, or composites, have attracted an ever growing interest from the scientific community, especially for thermal management or energy harvesting applications\cite{Chen2012535}, in the form of superlattices~\cite{Hu2012,Broido2004,Termentzidis2009,france14b}, core/shell nanowires~\cite{Yang2005,Nika2013,Blandre2015,Pavloudis2016}, and particulate nanocomposites~\cite{Poon2013,verdier2016,Huang2017,Tlili2017,Damart2015}. 
   
Specifically, the nanostructuration and the local disorder around nanoscale interfaces has proved to be an efficient way for tailoring thermal properties, and in particular reducing heat transport. A hierarchical phonon scattering has been proposed for drastically decreasing the thermal conductivity, through a combination of atomic-scale nanostructuring, endotaxial precipitation and scattering from mesoscale grain boundaries, which would affect all phonon wavelengths~\cite{Kanatzidis2014}. Besides thermal management, such composites have arisen as most promising for thermoelectric applications, where low thermal conductivity but good electrical properties are sought for optimizing the figure of merit $ZT=(S^2 \cdot \sigma \cdot T)/\kappa$, where $S$ is the Seebeck coefficient, $\sigma$ is the electrical conductivity, $T$ the temperature and $\kappa$ the thermal conductivity. 
Indeed, Guo and Huang~\cite{Guo2015} showed that thermal conductivity alloy limit can be achieved in a nanocomposite made of crystalline-nanoinclusions embedded in a crystalline matrix and suggested that depending on the fillers, improvement of the electrical properties might be achieved. According to them, the modification of the transport properties of phonons and electrons might lead to an increase of the ZT. 
In this context, Faleev \textit{et al.}~\cite{Faleev08} proposed that metallic nanoinclusions inside semiconductor matrix would enhance ZT, thanks to a drastic thermal conductivity reduction through different phonon scattering processes depending on the matrix, electron-phonon scattering for high doping and phonon-interface scattering for low doping levels. A major role of interface scattering was identified in a composite made of ZnO nanoinclusions embedded in a In$_2$O$_3$ matrix, proposed for high thermoelectric performance devices~\cite{wei2013}. Most of the studies concerning particle nanocomposites focused on crystalline or amorphous nanoinclusions embedded in crystalline matrix, where the purpose was to optimize the ZT of a material with good electric properties, by reducing its thermal conductivity through the introduction of scattering elements (nanoinclusions). 

The opposite strategy is possible, where inclusions are introduced in a bad thermal conductor with the purpose of improving its electric properties without significantly increasing the thermal conductivity. This line has been followed in a series of recent works, focused on composites made of silicon nanocrystallites incorporated in amorphous silicon or silicon dioxide~\cite{miura2015,Claudio2014,France-Lanord2014,Shiomi2016}. The main idea was here that the crystalline components can help to maintain or even rise the power factor ($P=S^2 \cdot \sigma$), while the amorphous matrix and the interfaces would still hinder heat transport.

In this work, we address the understanding of thermal transport in this kind of nanocomposites, where semiconductive crystalline particles are embedded in an insulating amorphous matrix, expected to exhibit a percolating behavior, and for this reason here called percolating nanocomposites. More specifically, we present a molecular dynamics investigation on a system which has recently been experimentally realized, made of GaN nanoclusters in an amorphous silica matrix~\cite{Borsella2001}. Some of us, using the right concentration ratios and annealing temperatures, could already model such composites, observing the cluster formation by means of Molecular Dynamics (MD) simulations~\cite{kioseoglou17}.  


In such a composite, we expect the electronic contribution to thermal conductivity to be negligible, thus this study provides most of the information on the nanoscale heat transfer. We will show here that, similarly to what is expected for electronic properties, a suprising percolating behavior is observed for phonon heat transport, with an enhanced efficiency that we associate to the most efficient phonon tunneling between iso-oriented nanoinclusions. As such, thermal transport in this kind of percolating composites is drastically different from the previously largely investigated nanocomposites where interface scattering plays a major role, so that the important parameters turn out to be different. 

\section{Simulation set up and structure modeling}

\subsection{Equilibrium Molecular Dynamics}

Molecular Dynamics is a powerful tool for calculating the thermal properties of nanostructured materials, because the length scale probed by the method is in the nanometer range and there are no assumptions made on the phonon dynamics except their classical nature. Here, the equilibrium molecular dynamic (EMD) approach is used. The principle of EMD relies on the fact that the regression of the thermal fluctuations of an internal variable, in our case the heat flux, obeys macroscopic laws. As a result, the time decay of the fluctuations of the heat flux is proportional to the thermal conductivity. In terms of mathematics, this is expressed by the Green-Kubo formula which states that the time integral of the heat flux autocorrelation function is proportional to the thermal conductivity tensor:

\begin{equation}\label{Green-Kubo formulae}
\lambda_{\alpha,\beta}=\frac{1}{Vk_{B}T^2}\int_0^\infty \langle J_\alpha(t) J_\beta (0) \rangle dt
\end{equation}

where V is the system's total volume and J$_i$ denotes the component of the heat flux vector along the direction $i$. The equilibrium method consists in computing the corresponding autocorrelation, which requires following the dynamics of a system over time scales a few times larger than the longest relaxation time present in the system. In the case of heat transfer in solids, the longest relaxation times correspond to long wavelength phonons (few nm) with a lifetime of about 100~ps. 

In all simulations, the three body Tersoff based interatomic potential has been used for all atomic combinations~\cite{Tersoff89,Brenner89}. The average temperature was set as 300~K by initializing atom velocities with a Maxwell-Boltzmann distribution and by relaxing the system with a NVT dynamic phase for a duration of 200 ps. Flux fluctuation correlations are calculated every 10 fs, while the thermal conductivity is estimated every 40 ps with equation~\ref{Green-Kubo formulae}. The thermal conductivity is averaged when the steady state regime is established. Due to CPU restrictions we repeated each simulation 5-10 times. Good statistics are obtained and the estimated uncertainty is less than 10\%.

\subsection{Modeling the structures}

The composite model, made of spherical crystalline GaN nanoinclusions embedded in an amorphous SiO$_2$ matrix, has been prepared as follows: 

(a) an initial amorphous SiO$_2$ bulk was prepared with the desired structural characteristics (radial distribution function, density) ~\cite{kioseoglou17}. The size of the SiO$_2$ box was 10.164~nm by 13.203~nm by 13.819~nm, with a total volume of 1854.44~nm$^3$.

(b) Depending on the size and spatial distribution of the GaN nanoinclusions to be incorporated in the SiO$_2$ matrix, corresponding spherical regions were removed from the amorphous bulk, thus creating nanoporous silica with spherical porous.

(c) Nanoclusters of crystalline GaN were prepared with volumes equal to the deleted volumes in the silica matrix (actually with radius smaller than the empty spaces in silica matrix by 0.1~\AA to avoid strains).
 
(d) The two structures (crystalline GaN clusters and nanoporous silica) were merged in one.

(e) A conjugated gradient relaxation procedure was finally applied to the composite system to achieve a relaxed structure. All systems studied here resulted energetically stable and no displacement, merging, fusion or alloying of the nanoinclusions was observed. This was counter-checked at room temperature on a couple of composite structures, with long runs. We expect movement and eventual merge of the nanoinclusions for temperatures higher than 1000~K as it has been observed in a previous work~\cite{kioseoglou17}.

Periodic boundary conditions are applied in all directions and all systems. The center of the nanoinclusion, when only one nanoinclusion is considered, was set at the center of the simulation box. In the case of more than one nanoinclusion per system, their centers were spatially homogeneously arranged.

\begin{figure}[h]
\centering
	\includegraphics[height=4cm]{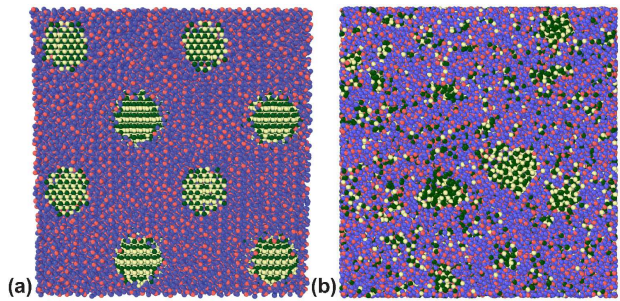}
	\vspace{25px}
	\caption{Two modeled systems with the same volume fraction of GaN nanoinclusions inside an amorphous silica matrix: (a) cross section of the system including 16 nanoinclusions with the same orientation (here only 8 of them are evident, as the other 8 are located in a non-visible plane), (b) ellipsoidal crystalline nanoinclusions after the formation procedure studied in a previous work\cite{kioseoglou17}. The cut-plane is made for a given depth of the simulation box and only a couple of nanoinclusions with different sizes can be seen. This is due to the specific cut that does not intersect all the inclusions at their maximum cross section.}\label{figure0a}
\end{figure}

Five different sets of nanocomposite structures have been created with this procedure. We summarize them in Table~\ref{table1}, where we also detail the name, the radius, the total surface area (in case of more than one nanoinclusion), the total volume fraction, the ratio of the surface/volume {\it and the calculated thermal conductivity}. They consist of the following systems:

(1) a structure containing 16 spherical crystalline GaN nanoinclusions with a radius of 1.5~nm, shown in figure~\ref{figure0a}-a. The spatial distribution of the nanoinclusions is uniform. This structure is created for comparison purposes with previously reported work~\cite{kioseoglou17}, aimed at assessing the importance of the nanoparticle shape for heat transport properties. Indeed, in that work, the GaN cluster formation in a silica matrix was studied, and  nanoclusters of an elliptical shape were obtained (figure~\ref{figure0a}-b), with some size dispersion. The aim here is to understand if the model system with monodisperse spherical particles is representative of the real one for investigating thermal transport. This configuration is labeled {\it 16clusters}~(see table~\ref{table1}).
(2) Seven systems with a single nanoinclusion, with different volume fractions, realized changing the radius of the GaN cluster from 1 to 5~nm. These structures are named {\it R$_1$-R$_7$} (table~\ref{table1}).
(3) Three systems with one, two and three nanoinclusions of different radius, but the same total interfacial area. These structures are called {\it S$_1$-S$_3$} (table~\ref{table1}). Note that the sample S$_1$ is the same as R$_5$.
(4) Three systems with two, three and four nanoinclusions, with different radius, but the same volume fraction. These structures are named {\it VF$_1$-VF$_3$} (table~\ref{table1}). It's important to specify here that in the S$_1$-S$_3$ and VF$_1$-VF$_3$ configurations, all GaN nanoinclusions have parallel crystallographic orientation.
(5) One system with two nanoinclusions and the same volume fraction as VF$_1$ sample but in this case the two nanoinclusions inside the box are misoriented by a relevant 60$^\circ$ in both x and y-directions. This structure is designed {\it arbitrary} in table~\ref{table1}.

%

\begin{table*}[h]
\small
\centering
\caption{Five different sets of nanocomposite structures of crystalline GaN nanoinclusions in a silica matrix have been modeled, each one identified by different color, which corresponds to the color of points in the figures with results. The label of all configurations, the number of nanoinclusions in the simulation box, the radius of nanoinclusions, their total surface area of interfaces (in case of more than one nanoinclusions), their total crystalline volume fraction and their ratio of surface to volume are given in this table.}
\label{table1}
\begin{tabular}{lcccccc}
\hline
%
%
sample name & $\sharp$ incl. & r (nm) & S (nm$^{-2}$) & VF (\%) & S/V (nm$^{-1}$) & TC (W/mK)\\ 
\hline
\textbf{16 nanoinclusions}  & 16 & 1.5 & 452.32 & 12.20 & 2 & 2.18 \\
\hline
\textbf{R$_1$}       & 1  & 1   & 12.57  & 0.23  & 3 & 1.23 \\
\textbf{R$_2$}       & 1  & 1.5 & 28.27  & 0.76  & 2 & 1.75 \\
\textbf{R$_3$}       & 1  & 2   & 50.27  & 1.81  & 1.5 & 3.40 \\
\textbf{R$_4$}       & 1  & 2.5 & 78.54  & 3.53  & 1.2 & 6.30  \\
\textbf{R$_5$}       & 1  & 3  & 113.10  & 6.10  & 1 & 10.76  \\
\textbf{R$_6$}       & 1  & 4  & 201.06  & 14.46 & 0.75 & 54.69 \\
\textbf{R$_7$}       & 1  & 5  & 314.16  & 28.23 & 0.6 & 98.01  \\
\hline
\textbf{S$_1$}       & 1  & 3  & 113.10  & 6.10  & 1 & 10.76  \\
\textbf{S$_2$}       & 2  & 2.12 & 113.10 & 2.16 & 1 & 6.48  \\
\textbf{S$_3$}       & 3  & 1.73 & 113.10 & 1.17 & 1 & 2.35 \\
\hline
\textbf{VF$_1$}      & 2  & 2.381 & 142.48 & 6.10 & 1.26 & 24.04 \\
\textbf{VF$_2$}      & 3  & 2.08  & 163.10 & 6.10 & 1.44 & 12.17 \\
\textbf{VF$_3$}      & 4  & 1.89  & 179.55 & 6.10 & 1.59 & 7.88 \\
\hline
\textbf{arbitrary}   & 2  & 2.381 & 142.48 & 6.10 & 1.26 & 14.76 \\
\hline
\end{tabular}
\end{table*}



\section{Results}

\subsection{Elliptical versus spherical nanoinclusions}\label{results1}

The thermal conductivity of bulk amorphous SiO$_2$ has been calculated with the EMD method and was found to be 1.14~W/mK, in good agreement with experimental findings (1.37~W/mK)~\cite{Shenogin2009}, theoretical calculations~\cite{Asegun2016,Jund1999} by GKMA (Green-Kubo Modal Analysis) method (1.32~W/mK) and MD (1.17~W/mK), taking into account that with classical dynamics, quantum corrections are not considered. 
As a first step, we have investigated thermal transport in two composites of 16 nanoinclusions, with a volume fraction of 12.2\%, one obtained during the formation of crystalline GaN nanocluster inside silica~\cite{kioseoglou17}, and characterized by polydisperse elliptical particles, and one with monodisperse spherical particles with the same average radius and with uniform spatial distribution. In the first case, the thermal conductivity has been found to be 1.97 $\pm$ 0.2 ~W/mK, while for the second composite system it was found equal to 2.18 $\pm$ 0.2 ~W/mK. These two results are very close and that proves that the second structure, which is much easier to be modeled, represents well the system with the formed GaN clusters. We cannot safely ascribe the 10\% difference to a weak effect of the nanoinclusions shape, as in fact the GaN clusters in the first sample are more defective than the spherical GaN crystalline inclusions modeled here, which could be the real reason for such discrepancy. Hereafter only spherical nanoinclusions are considered.

\subsection{Thermal conductivity versus volume fraction}\label{results2}

\begin{figure}[h]
\centering
	\includegraphics[height=6cm]{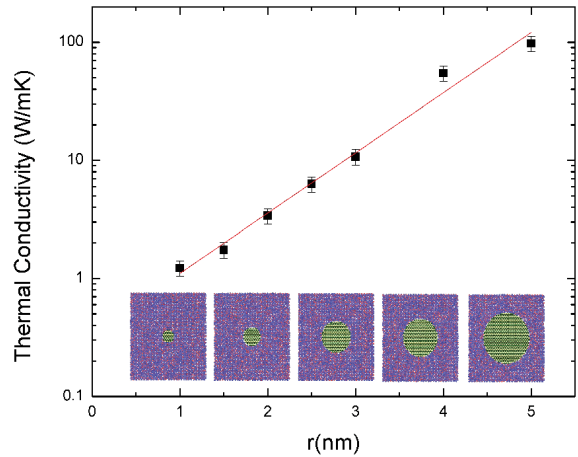}
	\caption{Computed TC with MD as a function of nanoinclusion radius }\label{figure2_a}
\end{figure}

\begin{figure}[h]
\centering
	\includegraphics[height=7cm]{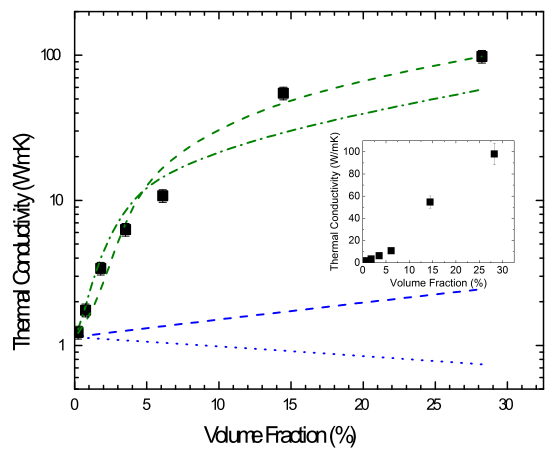}
	\caption{Computed (black points), EMA model (with and without R$_K$ blue dot line and blue dash line respectively) and EMA with percolation (dash-dot for normal set of parameters and dash for 2VF) TC as a function of volume fraction. In the inset the MD results are desplaced to show clearly the thresold of percolation.}\label{figure2_b}
\end{figure}

Initially the effect of the radius of crystalline GaN nanoinclusions on the system's thermal conductivity was studied. For this purpose, seven cases were examined, with the radius varying from 1 to 5~nm, as reported in Table~\ref{table1}. The thermal conductivity is depicted as a function of the nanoinclusion's radius in a logarithmic scale in Fig.~\ref{figure2_a}, revealing a surprising exponential dependence of the thermal conductivity on the nanoinclusions radius. When the nanoinclusion has a radius of only 2~nm, the composite displays almost the same thermal conductivity as the matrix, indicating a heat transport not perturbed by the presence of nanoinclusions. The opposite is true for the largest nanoinclusions, diameter of 10~nm, for which the thermal conductivity increases up to 100~W/mK. Changing the inclusion radius, while keeping the simulation box size constant, corresponds to changing the crystalline volume fraction. In order to get a better insight on the effect of the nanoinclusions on heat transport, we report thus the thermal conductivity as a function of the nanoinclusions volume fraction (VF) in Fig.~\ref{figure2_b}. Here a threshold-like behavior for the thermal conductivity is found. For volume fractions lower than $\approx 5$\% only a weak increase is observed, while for larger volume fractions the thermal conductivity dramatically increases, approaching the value for bulk GaN ($\approx160$~W/mK) already for volume fractions of about 30\%. We have tried to understand the observed dependence using the popular Effective Medium Approach (EMA), quite successful in describing macroscopic composites~\cite{Minnich2007}.The general formulation for spherical clusters is given by~\cite{Nan1997}:

\begin{equation}
\kappa_{total}^{EMA}=\kappa_{a-SiO_{2}}^{bulk} \frac{\kappa_{GaN}^{bulk}(1+2\alpha)+2\kappa_{a-SiO_{2}}^{bulk}+2VF[\kappa_{GaN}^{bulk}(1-\alpha)-\kappa_{a-SiO_{2}}^{bulk}]}{\kappa_{GaN}^{bulk}(1+2\alpha)+2\kappa_{a-SiO_{2}}^{bulk}-VF[\kappa_{GaN}^{bulk}(1-\alpha)-\kappa_{a-SiO_{2}}^{bulk}]}
\label{EMAeq}
\end{equation}

where $\kappa_{total}^{EMA}$ is the effective thermal conductivity of the whole system, $\kappa_{a-SiO_{2}}^{bulk}$ is the bulk thermal conductivity of the a-SiO$_2$ matrix, $\kappa_{GaN}^{bulk}$ is the bulk thermal conductivity of the crystalline GaN, while

\begin{equation}
\alpha=R_K^{a-SiO_2/GaN}\kappa_{a-SiO_{2}}^{bulk}/r
\label{EMAeq1}
\end{equation}

is a dimensionless parameter. The $r$ is the radius of the nanoinclusions and $R_K^{a-SiO_2/GaN}$ is the thermal boundary resistance between silica and GaN. In Fig.~\ref{figure2_b}, we report the thermal conductivities calculated with Eq.~\ref{EMAeq}, using the literature value for the thermal boundary resistance, $R_K^{a-SiO_2/GaN}=1.02~10^{-7}$ m$^2$K/W~\cite{Wang2015}, as well as neglecting it, which corresponds to putting the parameter $\alpha$ to 0. It can be seen that the full EMA clearly fails in reproducing our results: indeed, the calculated thermal conductivity surprisingly decreases with the volume fraction rather than increasing, indicating that the interface resistance should dominate the behavior for such small nanoinclusions. If we neglect the interfacial resistance, a more intuitive increase is found, but much weaker than the one obtained from our simulations. 

It is not new that the EMA fails in describing composites on the nano-scale, as it does not take into account the enhanced phonon interface scattering. Recently, Minnich~\cite{Minnich2007} has proposed a modified EMA, taking into account both phonon scattering in the matrix due to the nanoparticles and within the nanoparticle. For this, the nanoparticle thermal conductivity should be used, rather than the bulk value, as we have used instead here for GaN. Still, the introduction of such corrections could only decrease the EMA prediction, thus resulting in an even worse agreement with our data.

The surprising threshold behavior can be intuitively understood in terms of a percolating system. Indeed, we have an insulating matrix, in which we introduce conductive inclusions. Both for electric and thermal transport, it is expected that beyond a certain threshold volume fraction the inclusions create a percolating network, making it possible the establishment of a current or heat flow. This can be described by the following percolating EMA equation~\cite{Vaney2015}:

\begin{equation}
VF \frac{\kappa_{GaN}^{\frac{1}{t}}-\kappa_{total}^{\frac{1}{t}}}{\kappa_{GaN}^{\frac{1}{t}} + A \kappa_{total}^{\frac{1}{t}}}+(1-VF)\frac{\kappa_{SiO_2}^{\frac{1}{t}}-\kappa_{total}^{\frac{1}{t}}}{\kappa_{SiO_2}^{\frac{1}{t}} + A \kappa_{total}^{\frac{1}{t}}}
\label{percolationEq}
\end{equation}

with $t$  a parameter representing the asymmetry of the microstructure (in terms of inter-grain connectivity), and $A$ a function of the percolation threshold $VF_c$: $A=\frac{(1-VF_c)}{VF_c}$.
We report in Fig.~\ref{figure2_b} the percolating model best fit to our data, which corresponds to a percolation threshold of only $VF_c\approx 1.3$\% and a microstructure parameter $t=0.77$ (green dash-dot line). Beside the surprising result that the system percolates since very small volume fractions, the model is still unable to reproduce the large thermal conductivities observed at the largest volume fractions. Intuitively, it is like the effective volume fraction was larger than the real one. In order to test this idea, we have fitted the data with the same model, but using an effective volume fraction $VF_{eff}= b VF$, with $b$ a fitting variable (green dash line). The agreement with the simulation data is now amazingly good, as can be appreciated in the figure~\ref{figure2_b}. The resulting parameters are $VF_c=5$\%, $t=0.79$ and $B=2$.

\begin{figure}[h]
\centering
	\includegraphics[height=7cm]{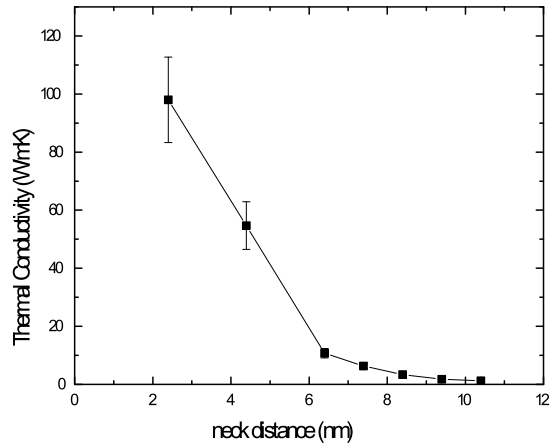}
	\caption{TC as a function of the amorphous neck distance between two nanoinclusions}\label{TC_neck_lin}
\end{figure}

How can we explain such behavior? A phonon tunneling - i.e. percolating mechanism - can explain the strong increase of the thermal conductivity with the volume fraction. Recently, Yan et al.~\cite{Yang2017} have showed that phonons can ballistically penetrate through a layer of amorphous SiO$_2$ of up to 5~nm thickness between two double silicon nanoribbons. If we look the behavior of the thermal conductivity as a function of the amorphous neck between nanoinclusions (figure~\ref{TC_neck_lin}), we find that it strongly increases for amorphous necks less than 6~nm, in a nice agreement with their findings. What about the larger effective volume fraction needed for describing the observed behavior with the standard percolating EMA? It suggests that the percolation efficiency is somehow enhanced here.

Once percolation can take place, the resulting thermal conductivity will depend on the transmission probability of phonons from one nanoinclusion to another, which is 1 if the particles are aligned, but lower if they are crystallographically misaligned. In our system, due to the periodic boundary conditions, all particles are obviously  aligned, and thus this transmission probability is 1 for all phonons. Still, the percolating model is a general treatment which  considers an isotropic matrix and isotropic - or isotropically distributed - inclusions. It should thus predict the behavior for randomly oriented particles, which would correspond to transmission functions lower than 1 and averaged over all possible misorientations. The prediction should thus result lower than the actual value in our case.

\begin{figure}[h]
\centering
	\includegraphics[height=4cm]{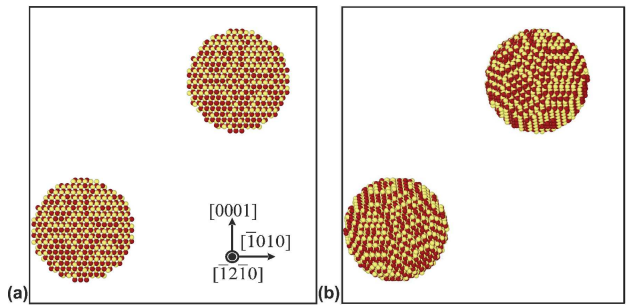}
	\caption{Two modeled systems with the same volume fraction of two GaN nanoinclusions inside an amorphous silica matrix: (a) the two nanoinclusions have the same relative orientation, (b) the two nanoinclusions have been rotated by 60$^\circ$ in both x and y directions.}\label{figure8}
\end{figure}

This is indeed the case, if the real volume fraction is considered. We can check the influence of the particle alignment on the calculated thermal conductivity by introducing a misalignment. To this purpose, we have prepared two systems with two nanoinclusions with $r=2.381$~nm, corresponding to $VF=6.1$\%: one where the two inclusions have the same orientation (system VF$_1$ in Table~\ref{table1}), and one where they are misaligned (system $arbitrary$ in Table~\ref{table1}, see Figure~\ref{figure8}).  The thermal conductivity decreases by 24.89~W/mK for the system with the two iso-oriented nanoinclusions to 14.76~W/mK for the system with the arbitrarily oriented nanoinclusions, loosing thus almost a factor of 2, and resulting more in agreement with the expected value, as predicted by the standard percolating model for such a volume fraction. This result nicely confirms the crucial role played by the nanoinclusions relative orientation in determining the thermal transport in composite materials when this latter is dominated by tunneling mechanisms. 
We can understand then the effective volume fraction as a way of taking into account the effectiveness of the particles iso-orientation for the percolation effect.
  
The dominating tunneling mechanism makes this composite different from many others already investigated in the literature, such as nanoporous systems~\cite{verdier2016} or SiGe nanocomposites~\cite{Minnich2007}, where the inclusions were less conductive and the dominant mechanism was interface scattering. There, it was possible to identify the leading parameter in determining heat transport, such as the volume fraction or neighbors distance or interface density. We can wonder if any of these parameters still plays an important role in our system. In the next section we thus discuss these issues.

\subsection{Thermal conductivity, number of inclusions, volume fraction and interface density}\label{results3}

\begin{figure}[h]
\centering
	\includegraphics[height=6cm]{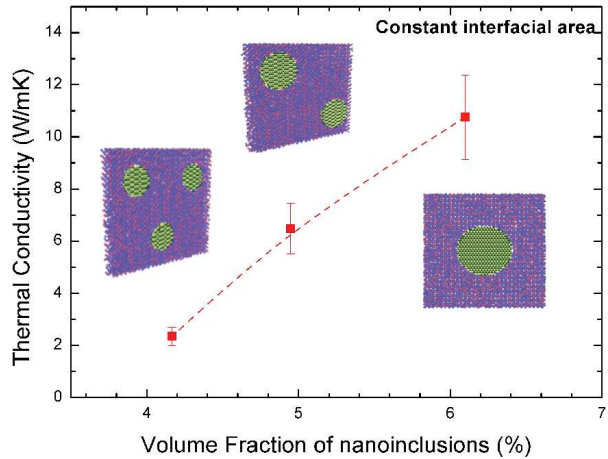}
	\caption{TC as a function of the Volume Fraction upon increasing the number of nanoinclusions, but for the same total interfacial area.}\label{figure2b}
\end{figure}

In this section, we investigate first the role of the number of crystalline inclusions, when the total interface is kept constant. For this, we consider as a reference the system with one nanoinclusion with radius 3~nm, which has interfacial surface equal to $113.1~~nm^2$, and we model two additional systems, with the same interfacial surface, but more than one inclusion. The behavior of the thermal conductivity as a function of the volume fraction is reported in Figure~\ref{figure2b}.

As expected, upon adding more nanoinclusions with smaller radius but same total interfacial area, the thermal conductivity decreases. Such a behavior has been observed long time ago in optics, where scattering in particulate media has been shown to be more efficient when the surface to volume ratio of the single particle increases (Mie's theory). This could be as well explained as a pure effect of the decreased volume fraction, suggesting thus the volume fraction to be the leading parameter. 
  
\begin{figure}[h]
\centering
	\includegraphics[height=6cm]{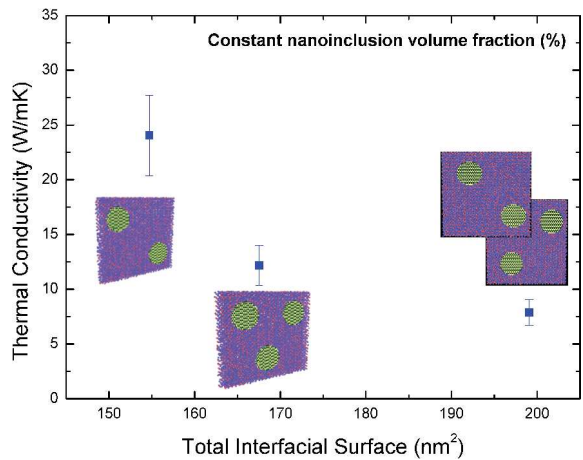}
	\caption{(Color online) TC as a function of the total interfacial area upon increasing the number of nanoinclusions at constant volume fraction (b). The facets of the insets are different each time to visualise in a more efficient way the spatial distribution of the nanoinclusions.}\label{figure3}
\end{figure}

%
%
In order to test this hypothesis, we have investigated another series of samples with multiple inclusions, prepared in such a way that the volume fraction is now constant, while the total interface changes. The reference value for the VF is always the same, as in the previous section, corresponding to a single nanoinclusion with a radius of 3~nm. In figure~\ref{figure3} the thermal conductivity as a function of the total interfacial area is depicted. It can be seen that a clear decrease with the interfacial area is observed, indicating thus that the volume fraction is not univocally leading heat transport. Putting together the constant interface and constant VF cases, we observe a reduction of thermal conductivity essentially related to the introduction of a larger number of smaller inclusions, which seems thus to be the key element.

\begin{figure}[h]
\centering
	\includegraphics[height=6cm]{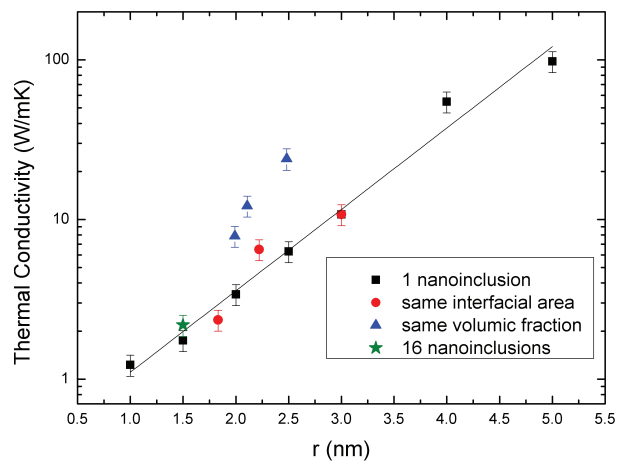}
	\caption{TC for all studied systems as a function of the radii of the nanoinclusions.}\label{figure4}
\end{figure}

This could suggest the existence of a leading role of the particles dimension, that we check plotting the thermal conductivity of all simulated systems as a function of the particles radius in Fig.~\ref{figure4}. While a majority of data follow the exponential behavior already found for the systems with 1 inclusion, some values do not align with the others: these are the constant VF cases.

\begin{figure}[h]
\centering
	\includegraphics[height=6cm]{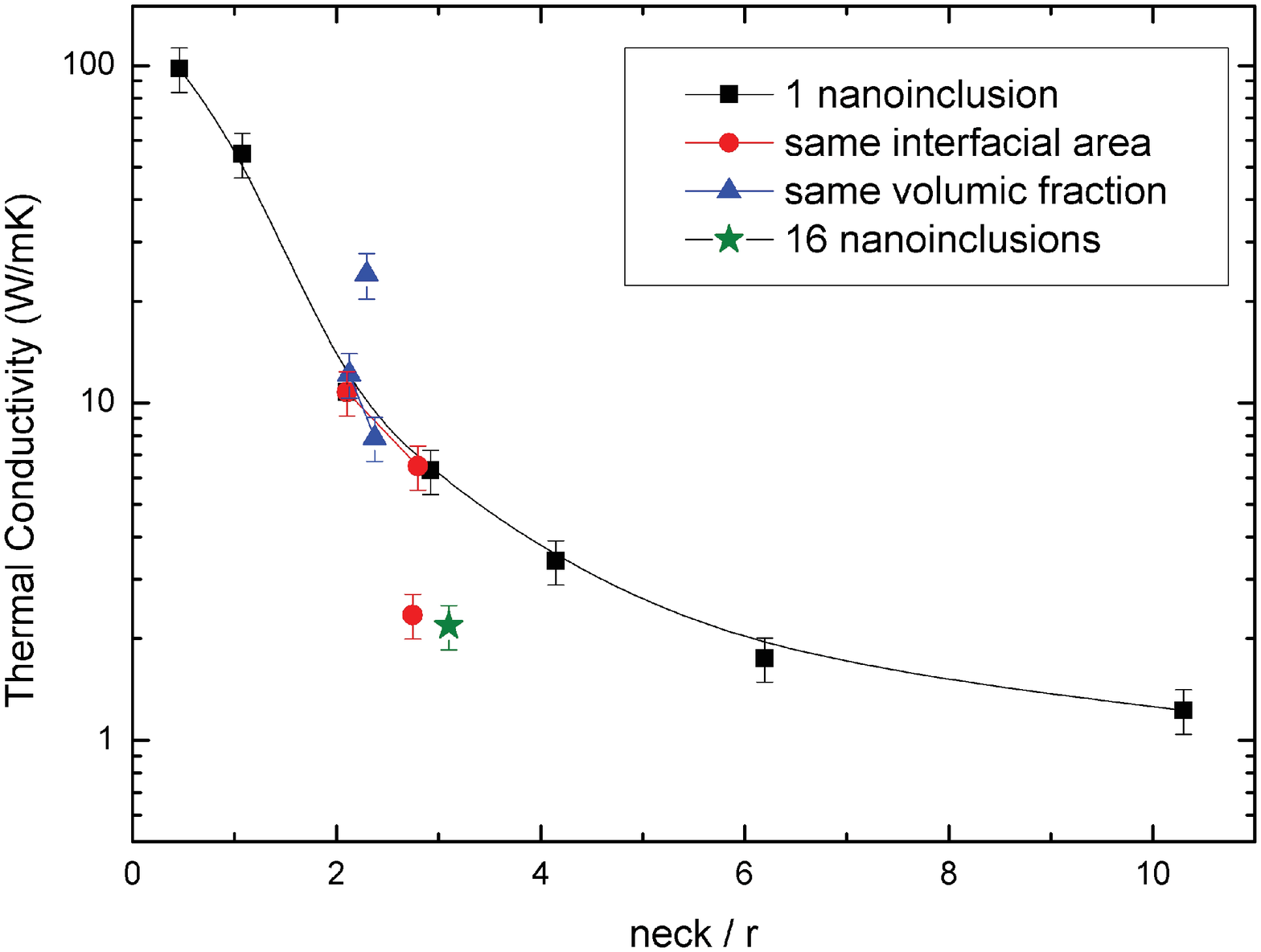}
	\caption{TC for all studied systems as a function of the ratio of the distance between two neighboring nanoinclusions (neck) divided by the radius of the nanoinclusions.}\label{figure5}
\end{figure}

Being in the case of a percolating system, it is clear that the inter-particle distance, the neck, must play an essential role. While a universal behavior cannot be found as a function of the neck, a better agreement arises when all data are reported as a function of the neck to radius ratio, as shown in Fig.~\ref{figure5}.

The overall thermal conductivity decreases in increasing this ratio which was expected as larger neck means larger path inside the amorphous matrix for phonons to cross between two crystalline nanoinclusions. Nevertheless there are still some points outside the main trend. These are the 16 nanoinclusions system, the three nanoinclusions system for constant total interfacial area and the two nanoinclusions for constant VF. For the first one (16 clusters), one could think that the system is too complex and there are too many parameters involved to be able to distinguish the predominant parameter which affects thermal transport. Concerning the other 2 systems, the argumentation is not so simple. A deeper analysis of the relative phonon mean free paths of silica and crystalline nanoinclusions should be conducted. It is likely that thermal transport is the result of a competition between nanoinclusion diameter, distance between two neighboring nanoinclusions, interfacial surface and volume fraction, each one playing its own role and dominating in some parameters ranges.


\section{Discussion-Conclusions}





In this work we have addressed a nanocomposite made of crystalline semiconductive nanoparticles embedded in an amorphous insulating matrix. Very few works can be found in literature on this kind of composites, but all of them highlight a leading role of the crystal/amorphous interface. 
Recently, Tanguy et al~\cite{Termentzidis_book_2017}, reported a molecular dynamics study of crystalline nanoinclusions inside an amorphous matrix in purely Si-like systems. They noticed that the vibrations of the matrix and of the nanoinclusions are decoupled for very few cases, but in any case these latter are different from the ones of an isolated cluster of the same size, clearly indicating a mutual effect between matrix and particles. Similarly, Damart et al~\cite{Damart2015} showed that the acoustic properties of such a nanocomposite are not a simple combination of the vibrations in the inclusions and in the amorphous matrix. More interestingly, they found that the presence of crystalline/amorphous interfaces can act as a low-pass filter for acoustic phonons, as they slow down the high energy components of a traveling wave packet. Another recent study involving amorphous/crystalline interfaces~\cite{France-Lanord2014} revealed that such interfaces perturb phonons up to a distance of 1~nm from them. Finally, experimental microscopic information on phonon dynamics in an amorphous/nanocrystalline composite has been recently reported~\cite{Tlili2017}, revealing no enhanced interface effect on phonons energies and lifetimes, which was ascribed to the lack of elastic contrast between the two composite constituents. 

In this work we have shown that the large contrast of thermal conductivity between the components, with the most conductive being the host particles, takes over the leading role in thermal transport, causing a percolating behavior, where phonon tunneling between the nanoinclusions is further enhanced by the iso-orientation of these latter. 
In such kind of composites, the interface scattering looses its efficiency, and is almost negligible. 
Still, a universal behavior of the thermal conductivity in all the systems here investigated as a function of the inclusions distance or volume fraction or amorphous neck, as could be expected if percolation was the only dominant mechanism, could not be found. \\
Our work indicates thus that interface scattering and phonon tunneling are in competition in determining thermal transport, likely dominating each in a certain range of parameters such as inclusions size, distance, total interface and volume fraction. Their effective role will also depend on the phonons involved in thermal transport, and their propagative or diffusive nature, phonon mean free path or diffusion time. 
Many questions thus remain open, asking for novel theoretical studies, going into the detailed behavior of the individual phonons, as well as for microscopic experimental investigations, able to guide and constraint the theoretical models.

\section*{Acknowledgements}
This work was supported by computational resources granted from the Greek Research \&  Technology Network (GRNET) in the National HPC facility `ARIS' under the project AMONADE (ID pr004002), from 'ERMIONE' cluster (IJL-LEMTA) and from IDRIS. The authors thank P. Ben-Abdallah (Laboratoire Charles Fabry, Institut d'Optique) and S. Gomes (CETHIL) for fruitful discussions.   

%


\bibliographystyle{unsrt}

\end{document}